\documentclass[10pt]{article}

\usepackage{comment} 
\usepackage{graphicx}
\usepackage{dcolumn}
\usepackage{bm}

\def \ccomma{\raise 2pt\hbox{,}} 
\def \D {\hbox{d}}

\def\barA{{\bar A}}
\def\tanphi{\psi}

\def \cotg {\mathop{\rm cotg}\nolimits}

\def \cs {\mathop{\rm cs}\nolimits}

\def \dn {\mathop{\rm dn}\nolimits}

\def \Halphen {\mathop{\rm H}\nolimits} 
\def \Halphen {\mathop{\raise 0.15pt\hbox{\rm H}}\nolimits} 
\def \ha      {\mathop{\raise 0.15pt\hbox{\rm h}}\nolimits} 
\def \Re  {\mathop{\rm Re}\nolimits}
\def \Im  {\mathop{\rm Im}\nolimits}

\def \mod#1{\vert #1 \vert}
\def \ReA {R}
\def \ImA {I}
\def \FIJ {h}   
\def \oms {\omega_0} 
\def \Ome {\Omega  } 

\def \kH   {k_2}       
\def \kD   {k_1}       

\def \at {a}      
\def \kt {k}      
\def \Kt {\kappa} 
\def \ax {\alpha} 
\def \kx {\lambda}
\def \Kx {K}      
\def \iKx {{K'}}  
\def \tone{{t_1}}
                              
\def \Jmx {m_x}       

\def \WP    {\mathcal{P}}
\def \Wpts {V}
\def \NDLR#1 {\hfill\break\noindent NDLR \textbf{#1}\hfill\break\noindent}
 \def \url#1 {#1}

\title{Explicit breather solution of the nonlinear Schr\"odinger equation}

\author{Robert Conte
\\ Universit\'e Paris-Saclay, ENS Paris-Saclay, CNRS, 
\\ Centre Borelli, F-91190 Gif-sur-Yvette, France
\\ Department of Mathematics, The University of Hong Kong, 
\\ Pokfulam Road, Hong Kong
\\ E-mail Robert.Conte@cea.fr
\\ ORCID https://orcid.org/0000-0002-1840-5095
}

\date{April 12, 2021}

\begin{document}		
									
\maketitle

\abstract{
We present a one-line closed form expression for 
the three-parameter
breather of the nonlinear Schr\"odinger equation.
This provides an analytic proof of the time period doubling
observed in experiments.
The experimental check
that some pulses generated in optical fibers are indeed such generalized breathers
will be drastically simplified.
}

\textit{Keywords}:
modulational instability, 
nonlinear Schr\"odinger equation,
nonlinear optics,
breather,
exact solutions.

PACS									
02.30.Hq, 
02.30.Jr, 
03.75.Kk, 
03.75.Lm, 
42.65.J,  
42.65.-k	

\tableofcontents
\section{Introduction}
\label{sectionNLS-Breather}

The complex amplitude of many nonlinear media displays two generic features.
The first one is to obey an evolution equation (first order in the time variable $t$)
with a (linear) dispersion term (second order in the space variable $x$)
and the simplest nonlinearity preserving the phase invariance of $A$,
\begin{eqnarray}
& & {\hskip -4.0truemm}
i A_t + p A_{xx} + q \mod{A}^2 A =0,\
p q \not=0,\ p,q \hbox{ real}. 
\label{eqNLS}
\end{eqnarray}
The second feature, observed in the ``focusing'' r\'egime ($p q >0$) 
of this nonlinear Schr\"odinger equation (NLS),
is the ``modulational instability'' (MI) \cite{Bespalov-Talanov},
also known as Benjamin-Feir instability:
an initial plane wave grows exponentially, then saturates and decreases 
to its original state, with only a shift of its phase.
This MI has enormous applications, which we now recall.

In the ocean, deep water waves are suitably described by the focusing NLS \cite{Zakharov1968},
where one observes ``bright'' solitons.
Sailors have also reported the sudden occurence of huge waves (``freak'' or ``rogue'' waves)
which disappear as quickly as they appeared,
and these solutions of very high amplitude and energy can also be described by NLS \cite{AAT}.
However, experimental setups able to reproduce this rare observation are quite difficult.
As to the  ``defocusing'' r\'egime ($p q <0$), it is more adapted to shallow water waves,
where only ``dark'' solitons occur.

The situation is quite different in Bose-Einstein condensation (BEC),
where the wave function of the condensate obeys the Gross-Pitaevskii equation,
a three-dimensional analogue of NLS.
It has been proven analytically \cite{Konotop_Salerno.BEC-MI.2002} that MI
is the mechanism which generates wave functions of soliton type
in a Bose-Einstein condensate, a prediction confirmed 
by the experimental observation \cite{Everitt-et-al.BEC-MI.2017} 
of MI in a cigar-shaped BEC.

But nowadays the main playground of MI 
no more water waves nor even BEC but nonlinear optics, for two reasons.
The first one is the huge recent progress in manufacturing optical fibers
with prescribed physical properties (refractive index, etc),
making experiments easier, cheaper and easily reproducible.
The second reason is more fundamental:
as opposed to a three-dimensional BEC,
a fiber is quasi one-dimensional 
and thus well described by the NLS,
$t$ being the propagation distance and 
$x$ the transverse coordinate.
For instance, one has succeeded \cite{SRKJ,Yeom-Eggleton} 
to generate rogue waves in optical fibers,
an achievement with potentially important industrial applications.
Nonlinear optics has become an excellent field to perform an experimental check 
of the beautiful analytic description of MI, which we first recall.

Indeed, 
the later stages of MI can be computed exactly,
resulting in a two-parameter\footnote{%
The scaling invariance $(x,t,A) \to (k x, k^2 t, k A)$ of NLS
reduces this number by one.} 
bright soliton localized in space and periodic in time,
whose asymptotic behaviour as $\mod{x} \to + \infty$ is the plane wave 
$e^{-i \oms t}$, see (\ref{eq-NLS-Kuznetsov}) below.
This achievement of Kuznetsov \cite{Kuznetsov-NLS} was 
obtained in plasma physics where the Langmuir waves are appropriately described by the focusing NLS.
Changing the sign of one parameter converts this soliton to 
another quite important physical solution, localized in time and periodic in space,
known as the Akhmediev breather \cite{AK},
see (\ref{eq-NLS-Akhmediev}) below.
Finally, using a quite simple Ansatz,
Akhmediev, Eleonskii and Kulagin \cite{AEK}
extrapolated the Kuznetsov soliton to a three-parameter breather solution,
in which $A e^{i \oms t}$ is elliptic\footnote{
We never use the ambiguous term ``periodic'' for elliptic solutions,
but always 
either ``doubly periodic'' alias ``elliptic'' 
(example: Jacobi $\dn$, Weierstrass $\wp$),
or ``quasi-doubly periodic'' alias ``quasi-elliptic''
alias ``elliptic of the second kind'' in Hermite's terminology
\cite[tome I p.~227, tome II p.~506]{HalphenTraite}
(example: the solution $\Halphen(t,a)$ of Lam\'e equation (\ref{eqLame1})).}
in $x$ 
and quasi-elliptic in $t$.


In a recent experiment \cite{PhLAM.OL.2020} with an optical fiber,
this breather was observed
by matching the three arbitrary parameters with experimental data,
showing a ``good'' agreement, however during only two quasi-periods of $t$.
The difficulty did not arise from the sophisticated experimental setup
but from ``the complexity of this class of solutions'' \cite{PhLAM.PRA.2020}.
Indeed, its current analytic representation \cite[(3), (22), (24)--(25)]{AEK}
does not clearly separates
the elliptic dependence on $x$ and 
quasi-elliptic dependence on $t$,
despite several later attempts
\cite{AA-NLS-breather} 
\cite{MP-JMP} 
\cite{MP-PRA}
\cite{Chow2002},
forcing the authors to expand the amplitude in Fourier series of $x$
and to retain only the first two coefficients. 

In this article, we provide a one-line closed form expression for this
three-parameter solution, see Eq.~(\ref{eq-NLS-generic-ell}),
and perform a full classification of the solutions of the Ansatz of Ref.~\cite{AEK},
thus uncovering a new solution, Eq.~(\ref{eq-NLS-trigell}),
elliptic in $x$ and trigonometric in $t$, together with its degeneracy.
The present three-parameter closed form 
makes it possible to check the agreement on a much larger number of 
quasi-periods of $t$,
and therefore to determine more accurately 
the nonlinear range of validity of MI as sketched in \cite{PhLAM.PRA.2020}.
Another puzzling phenomenon observed in Ref.~\cite{PhLAM.OL.2020},
namely a time period twice the one expected,
is naturally explained by our three-parameter solution.

\section{The generic solution}
\label{sectionNLS-Breather-new-notation-AEK}

Ref.~\cite{AK} assumes 
a constraint between $A$ and $\barA$,
defined by three real functions $\varphi(t)$, $\delta(t)$
and $Q(x,t)$,
\begin{eqnarray}
& & {\hskip -9.0truemm}
\left\lbrace
\begin{array}{l}
\displaystyle{
 \sin \varphi(t) \Re(A) - \cos \varphi(t) \Im(A) + \delta(t)=0,\
}\\ \displaystyle{
 \cos \varphi(t) \Re(A) + \sin \varphi(t) \Im(A) - Q(x,t)=0.
}
\end{array}
\right.
\label{eqAK-constraint}
\end{eqnarray}
Since $A=(Q/\delta+i) (\delta e^{i \varphi})$ is single-valued \cite{CCT,CMBook},
both terms $Q/\delta$ and $\delta e^{i \varphi}$
are single-valued,
while $Q$, $\delta$ and $e^{i \varphi}$ may be multivalued.
Because of the absence of methods to handle multivaluedness,
the strategy is therefore to only consider 
$\delta e^{i \varphi}$, its complex conjugate and $Q/\delta$.

\textit{Remark}.
The real and imaginary parts of $A$ are,
\begin{eqnarray}
& & 
A =  [Q-\delta \tanphi +i (Q \tanphi + \delta)]/\sqrt{1+\tanphi^2},
\tanphi=\tan\varphi.
\end{eqnarray}

Let us first recall the result of \cite{AEK},
then proceed to the explicit dependence on $x$ and $t$.
By elimination of $A$, the system to be solved is made of
two coupled real PDEs for $Q(x,t)$ \cite[Eqs.~(4)--(5)]{AEK},
\begin{eqnarray}
& & {\hskip -9.0truemm}
\left\lbrace
\begin{array}{l}
\displaystyle{
Q_t + q \delta Q^2 - \varphi' \delta + q \delta^3=0,\
}\\ \displaystyle{
p Q_{xx} + q Q^3 + (q \delta^2-\varphi') Q - \delta'=0,\
}
\end{array}
\right.
\label{eqsystemQ}
\end{eqnarray}
and the second equation admits the first integral $\FIJ(t)$,
\begin{eqnarray}
& &
(Q_x\not=0)\ : 
\FIJ=p {Q_{x}}^2 + q Q^4/2 
+ (q \delta^2-\varphi') Q^2 - 2 \delta' Q.   
\label{systemQfirstintegral}
\end{eqnarray}
The integrability 
of (\ref{eqsystemQ})${}_1$ and (\ref{systemQfirstintegral}) 
defines the ODEs, 
\begin{eqnarray}
& & {\hskip -9.0truemm}
\left\lbrace
\begin{array}{l}
\displaystyle{
\varphi'' + 4 q \delta \delta'=0,\
\FIJ' + 2 \delta \delta' \varphi' - 2 q \delta^3 \delta'=0,\
}\\ \displaystyle{
\delta'' + \delta {\varphi'}^2 - 2 q \delta^3 \varphi'  + 2 q \delta \FIJ + q^2 \delta^5=0.
}
\end{array}
\right.
\label{eqsystemt}
\end{eqnarray}
This system admits three real first integrals $\oms, \kD, \kH$, 
\begin{eqnarray}
& & {\hskip -2.0truemm}
\left\lbrace
\begin{array}{l}
\displaystyle{
q \delta^2=2 z,\
\varphi'=- 4 z -\oms,\
q \FIJ=2(3 z^2+  \oms z + \kH),\
}\\ \displaystyle{	
(z'\not=0):\ {z'}^2= - 4 (4 z + \oms)^2 z^2 - 16 \kH z^2 + 4 \kD z,
}
\end{array}
\right.
\label{eqsystemtsol}
\end{eqnarray}
characterized by the three nonzero roots of $z'$,
\begin{eqnarray}
& & 
\left\lbrace
\begin{array}{l}
\displaystyle{
\oms=-2(z_1+z_2+z_3),\
\kD=16 z_1 z_2 z_3,\
}\\ \displaystyle{	
\kH=(z_1+z_2+z_3)^2-2(z_1^2+z_2^2+z_3^2).
}
\end{array}
\right.
\label{eqrktozj}
\end{eqnarray}

In the generic case $Q_x z \kD \not=0$
(nongeneric cases are detailed in section \ref{sectionTrigoSol}), 
the product $\delta^2$ 
is an elliptic function \cite[Eq.~(13)]{AEK} which
in the notation of Weierstrass\footnote{To convert to the notation of Jacobi,
see \cite[\S 18.9.11, 18.10.8]{AbramowitzStegun}.}
takes the quite simple form ($i a$ is real),
\begin{eqnarray}
& & 
(\kD\not=0)
\left\lbrace
\begin{array}{l}
\displaystyle{
z=\frac{\kD}{\wp(t)-\wp(a)},\                      
\wp(a)=-\frac{\oms^2 + 4 \kH}{3}\ccomma\
\wp'(a)=-8 i \kD,\ 
}\\ \displaystyle{ 
g_2=(4/3) \left\lbrack (\oms^2+ 4 \kH)^2 + 24 \kD \oms \right\rbrack,\
}\\ \displaystyle{
g_3=(8/27) \left\lbrack (\oms^2+ 4 \kH)^3 + 36 \kD (\oms(\oms^2 + 4 \kH) + 6 \kD) \right\rbrack,\
}\\ \displaystyle{
\Delta^{(t)} \equiv
g_2^3-27 g_3^2 = -2^{12} \kD^2 
}\\ \displaystyle{
\phantom{1} \times
\left\lbrack 16 \kH^3 + 8 \oms^2 \kH^2 + \oms^4 \kH +36 \oms \kH \kD + \oms^3 \kD +27 \kD^2\right\rbrack.
}
\end{array}
\right.
\label{eq-NLS-breather-t-elliptic}
\end{eqnarray}

Let us next determine simultaneously $\delta e^{i \varphi}$ and  $\delta e^{-i \varphi}$,
not by the multivalued quadrature $\int \varphi' \D t$ as usually done, 
but as the two complex conjugate solutions of a real second order ODE.
The phase invariance of NLS only allowing the contribution of $\varphi'$, 
not of $\varphi$,
by elimination of $z$
one easily obtains the Lam\'e equation of index $n=1$,
\begin{eqnarray} 
& & 
\left(\frac{\D^2}{\D t^2} -(2 \wp(t)+\wp(a))\right)\left(\delta^{-1} e^{\mp i (\varphi+\omega_0 t)}\right)=0. 
\label{eqLame1}
\end{eqnarray}
Its two independent solutions are generically,
\begin{eqnarray}
& & {\hskip -5.0 truemm}
\delta^{-1} e^{\mp i \varphi}
=       \sqrt{-q/\kD} e^{\displaystyle \pm i \oms t}\Halphen(t,\pm a),\
\label{eq-NLS-breather-delta-times-ephi}
\end{eqnarray}
with the definition 
\cite[tome II p.~506]{HalphenTraite}, 
\begin{eqnarray}
 & & \Halphen(t,a)=e^{\displaystyle -\zeta(a) t}\sigma(t+a)/(\sigma(a)\sigma(t)).
\label{eq-Element-simple}
\end{eqnarray}

At this point, Ref.~\cite{AEK} chooses to integrate 
the $x$-elliptic ODE (\ref{systemQfirstintegral})
with $t$-dependent coefficients.
It is more efficient to integrate the $t$-Riccati ODE (\ref{eqsystemQ})${}_1$
with $x$-independent coefficients,
and this will allow us to uncover a new solution, Eq.~(\ref{eq-NLS-trigell}).
Indeed,
an affine transformation on $Q(x,t)$ maps the equation 
(\ref{eqsystemQ})${}_1$ to
a canonical Riccati equation,
\begin{eqnarray}
& & {\hskip -2.0 truemm}
\left\lbrace
\begin{array}{l}
\displaystyle{
 (z\not=0):\ Q(x,t)/\delta(t) =y(x,t)/(2 z) + z'/(8 z^2)\ccomma\
}\\ \displaystyle{
\partial_t y + y^2 - (3/4)\wp(t)=0,\ 
}
\end{array}
\right.
\label{eqRiccatiy}
\end{eqnarray}
equivalent to a particular Lam\'e equation of index $n=1/2$,
whose solution is 
\cite[\S 20 p.~104]{Halphen-Lame-half} 
\cite[tome II p.~482]{HalphenTraite},
\begin{eqnarray}
 & & y=\partial_t \log \left\lbrack{\wp'(t/2)^{-1/2}} \left(4 \sqrt{\kD} F(x) + \wp(t/2) - \wp(a)\right) \right\rbrack.
\nonumber
\end{eqnarray}
The real-valued function $\sqrt{\kD} F(x)$ is defined by, 
\begin{eqnarray}
& & {\hskip -7.0truemm}
\begin{array}{l}
\displaystyle{
p {F'}^2 + P(F)=0,\ P(F)\equiv F^4 + \oms F^2 - 2 \sqrt{\kD} F - \kH,
}
\end{array}
\label{eq-NLS-breather-x-elliptic}
\end{eqnarray}
and evaluates to (all $\sqrt{}$ signs are allowed),
\begin{eqnarray}
& & {\hskip -2.0truemm}
\left\lbrace
\begin{array}{l}
\displaystyle{
F=\sqrt{z_1}+\sqrt{z_2}+\sqrt{z_3}
}\\ \displaystyle{\phantom{F=}
-\frac{2}{p}\frac{(\sqrt{z_2}+\sqrt{z_3})(\sqrt{z_3}+\sqrt{z_1})(\sqrt{z_1}+\sqrt{z_2})}
                 {\wp(x,G_2,G_3)- \wp(b,G_2,G_3)}\ccomma\
}\\ \displaystyle{
\wp(b)=-\frac{z_1+z_2+z_3+3(\sqrt{z_2}\sqrt{z_3}+\sqrt{z_3}\sqrt{z_1}+\sqrt{z_1}\sqrt{z_2})}{3 p}\ccomma\
}\\ \displaystyle{
\sqrt{\kD}=4 \sqrt{z_1}\sqrt{z_2}\sqrt{z_3},\
}\\ \displaystyle{
G_2=(\oms^2 - 12 \kH)/(12 p^2),
G_3=(\oms^3 + 36 \oms \kH + 54 \kD)/(6 p)^3,
}\\ \displaystyle{
\Delta^{(x)} \equiv G_2^3-27 G_3^2 
= 2^{-16} p^{-6} \kD^{-2} \Delta^{(t)}.
}
\end{array}
\right.
\label{eq-NLS-breather-x-elliptic-sol-F}
\end{eqnarray}

To summarize, the complex amplitude is, 
\begin{eqnarray}
& & 
\left\lbrace
\begin{array}{l}
\displaystyle{
%
A=
 \left\lbrace \frac{16 \sqrt{\kD}}{\wp'(t/2,g_2,g_3)} 
 \left\lbrack \frac{P(\Wpts)}{F(x) -\Wpts(t)} + \frac{\D P(\Wpts)}{4 \D \Wpts}\right\rbrack
 +i \right\rbrace  
}\\ \displaystyle{
\phantom{1234} \times
 \sqrt{-\kD/q} e^{\displaystyle -i \oms t}/\Halphen(t,a),\
}\\ \displaystyle{
\Wpts(t)=(\wp(a,g_2,g_3)-\wp(t/2,g_2,g_3))/(4 \sqrt{\kD}),
}
\end{array}
\right.
\label{eq-NLS-generic-ell}
\end{eqnarray}
with 
$P(\Wpts(t))$ and $F(x)$ defined in (\ref{eq-NLS-breather-x-elliptic}) and 
                                    (\ref{eq-NLS-breather-x-elliptic-sol-F}), 
$a$                              in (\ref{eq-NLS-breather-t-elliptic}), and
$\Halphen(t,a)$                  in (\ref{eq-Element-simple}),
and its complex conjugate results from the change $(i,a) \to (-i,-a)$.

This 
amplitude (\ref{eq-NLS-generic-ell})
depends on three arbitrary real constants
$\oms, \kD, \kH$
and is elliptic in $x$.
The ratio two between the $t$'s in $\wp(t/2)$ and in $\Halphen(t,a)$
makes the quasi-$t$-periods of $A(x,t) e^{i \oms t}$ 
twice the periods of $\wp(*,g_2,g_3)$,
thus proving the period doubling observation \cite{PhLAM.OL.2020}.

\textit{Remarks}.
\begin{enumerate}

\item
The generality of (\ref{eq-NLS-generic-ell}) is worth being emphasized.
This unique formula (the advantage of Weierstrass notation) covers 
both signs of the discriminant:
$\Delta^{(t)}<0$ (``B-type'' solutions, one   nonzero real $z_j$),
$\Delta^{(t)}>0$ (``A-type'' solutions, three nonzero real $z_j$),
it involves no multivalued expression
and even applies to both NLS r\'egimes (focusing, defocusing).

\item
The argument doubling formula 
$\sigma(2 y)=-\wp'(y) \sigma^4(y)$
\cite[18.4.8]{AbramowitzStegun}
allows one to express (\ref{eq-NLS-generic-ell})
with the unique argument $t/2$, i.e.~$t$
because of the homogeneity of $\wp$.

\item
To be physically admissible, 
the amplitude (\ref{eq-NLS-generic-ell}) must obey two constraints.
The first one $\delta^2(t)>0$
is: 
$p z_3>0$ \cite[p.~811]{AEK} and 
$p z_1,p z_2$ positive or complex conjugate, 
with bounds $0<p z(t)<\hbox{the smallest positive } p z_j$.
The second one $Q/\delta$ real, 
which was 
painful to implement \cite[p.~811]{AEK},
is equivalent to $\sqrt{\kD} F(x)$ real,
i.e.~
the transposition to the four zeroes $\sqrt{z_1}+\sqrt{z_2}+\sqrt{z_3}$ of $F'(x)$
of the constraints on the zeroes $0,z_1,z_2,z_3$ of $z'(t)$.
The bounded solutions of this focusing r\'egime 
result from (\ref{eq-NLS-generic-ell})
by applying to the origins of $x$ and $t$ 
a shift of either zero or a nonreal half-period,
depending on the common sign of the two discri\-minants $\Delta^{(x)}$, $\Delta^{(t)}$,
see formulae \cite[16.8,18.4.1]{AbramowitzStegun}.



\end{enumerate}

\section{Nongeneric solutions}
\label{sectionTrigoSol}

They are defined by
either 
$Q_{x}=0$ (inexistence of 
   $\FIJ(t)$) 
	or
$z'(t)=0$ (inexistence of 
   $k_1$) 
	or
$z(t)=0$ (undefined link (\ref{eqRiccatiy}) between $Q$ and $y$)
  or
$\kD=0$ (independence of (\ref{eqRiccatiy}) on $z$)
  or
$\wp'(a)=0$ (linear dependence of the two solutions (\ref{eq-NLS-breather-delta-times-ephi})
             of (\ref{eqLame1}))
  or
$\Delta^{(t)}=0$ (degeneracy of elliptic functions
     to either trigonometric functions or rational functions).
Because Eq.~(\ref{eqRiccatiy}) was not considered in \cite{AEK},
the nongeneric case $\kD=0$ will yield the new solution Eq.~(\ref{eq-NLS-trigell}).

\subsection{Degeneracies of the generic solution}
\label{sectionDegeneracy-generic}


They are characterized by 
$Q_x \delta' \kD \not=0$,
$\wp'(a) \Delta^{(t)}=0$.

When $\wp'(a)=0$,
then $a$ is a purely imaginary half-period $\omega'$,
the multipliers\footnote{
Under addition of anyone of the two periods,
a quasi-elliptic function is multiplied by a constant factor,
the multiplier.}
of $\Halphen(t,\omega')$ are $(-1,1)$
but  $\kD$ is zero, which is forbidden.
Fortunately,
the form invariance of the ODE for $\wp$
by halving one period
changes $\wp'(a)$ to $\wp'(2 a)\equiv 8 i (\kD+\oms \kH)$,
now allowed to vanish.
This ``Landen transformation'' 
\cite[p.~39]{Kiepert1873-wp-Landen}
\cite[16.14.2]{AbramowitzStegun}
%
\begin{eqnarray}
& &  {\hskip -4.0truemm}
\left\lbrace
\begin{array}{l}
\displaystyle{
\wp(t,     g_2,     g_3)\equiv\wp(t|\omega, \omega') \to 
\WP(t,\gamma_2,\gamma_3)\equiv\WP(t|\omega,2\omega'),\    
}\\ \displaystyle{
%
%
 \wp(t)=\WP(t)+(e_2-e_1)(e_3-e_1)/(\WP(t) - e_1),\
}\\ \displaystyle{
{\wp'}^2= 4 (\wp-e_1) (\wp-e_2) (\wp-e_3),\ 
e_1=(8 \kH - \oms^2)/3,\
}\\ \displaystyle{
{\WP'}^2= 4 (\WP+2 e_1) (\WP-\varepsilon_2) (\WP-\varepsilon_3),\
}\\ \displaystyle{
g_2=-4 \gamma_2 + 60 e_1^2,\
g_3= 8 \gamma_3 + 56 e_1^3,\   
}
\end{array}
\right.
\label{eq-Landen-wp-ascending}
\end{eqnarray}
makes both multipliers unity
(i.e.~$\Halphen$ elliptic),
yielding Jacobi functions as solutions to the Lam\'e ODE (\ref{eqLame1}),
\begin{eqnarray}
& &  {\hskip -6.0truemm}
\sqrt{\WP(t)-\varepsilon_2},\ \sqrt{\WP(t)-\varepsilon_3}.
\end{eqnarray}
This leads to the two 
elliptic breathers 
in an algorithmic way,
instead of the kind of magic derivation of Ref.~\cite{AEK},
and the notation of Halphen
\begin{eqnarray}
& &  {\hskip -2.0truemm}
\begin{array}{l}
\displaystyle{
\ha_ a    (x)=\sqrt{\wp(x,G_2,G_3)-E_ a},
\ha_\alpha(t)=\sqrt{\WP(t,\gamma_2,\gamma_3)-\varepsilon_\alpha},
}\\ \displaystyle{
{\wp'(x)}^2= 4 \wp^3 - G_2 \wp - G_3=4(\wp-E_a)(\wp-E_b)(\wp-E_c),
}
\end{array}
\label{eqWeierstrass}
\end{eqnarray}
allows one to unify them in a 
very symmetric expression.
Characterized by the relation $p z_1+p z_2=p z_3>0$
between the three roots of ${z'}^2$ in (\ref{eqsystemtsol})${}_3$,
their two singlevalued parts,
\begin{eqnarray}
& &  {\hskip -6.0truemm}
\left\lbrace
\begin{array}{l}
\displaystyle{
%
%
\delta e^{i \varphi}=\left(-\frac{\oms}{q}\right)^{1/2} \oms                  
\frac{\ha_1(t)e^{-i \oms t}} {\ha_1(t) \ha_3(t) + i \oms \ha_2(t)}\ccomma\
}\\ \displaystyle{
%
\frac{Q}{\delta}
=\frac{-\oms^2 \mu \ha_2(t) \ha_2(x)+\ha_1^2(t) \ha_3(t) \ha_3(x)}  
{\oms \ha_1(t)[\mu \ha_3(t) \ha_2(x)+\ha_2(t)            \ha_3(x)]}\ccomma\
}\\ \displaystyle{
\kH=(\mu^2-1)^2 \oms^2/2,\
\kD=-\oms \kH,
}
\end{array}
\right.
\end{eqnarray}
yield the amplitude 
\cite[Eq.~(18)]{AK}
\cite[Eqs.~(45), (59)]{AEK}
\begin{eqnarray}
& & \left\lbrace 
\begin{array}{ll}
\displaystyle{
%
%
%
A
=\left(-\frac{\oms}{q}\right)^{1/2}(\mu^2-1)                    
 \frac{\ha_\alpha(t)\ha_c(x) + i \mu \oms          \ha_b(x)}
      {\ha_\beta (t)\ha_c(x) +   \mu  \ha_\gamma(t)\ha_b(x)}
 e^{-i \oms t}
}\\ \displaystyle{
 \frac{\varepsilon_\beta -\varepsilon_\alpha}{(\mu^2-1)^2}
=\frac{\varepsilon_\gamma-\varepsilon_\beta }{-1}
=\frac{\varepsilon_\alpha-\varepsilon_\gamma}{\mu^2(2-\mu^2)}=\oms^2,\
}\\ \displaystyle{
 \frac{E_b-E_c}{2 (\mu^2-1)}
=\frac{E_c-E_b}{-\mu^2}
=\frac{E_a-E_c}{2-\mu^2}=\frac{\oms^2}{2 p}\ccomma\
}
\end{array}
\right.
\label{eqbielliptic_continuous_Halphen}
\label{eq-NLS-breather-AK}
\end{eqnarray}
with $(\alpha,\beta,\gamma)$ and $(a,b,c)$ two independent permutations of
$(1,2,3)$.
Its two arbitrary constants are $(\oms,\mu)$.
The conversion to Jacobi notation \cite[Appendix B]{CCX}
yields the two types A ($\Delta^{(t)}>0$) and B ($\Delta^{(t)}<0$).

Next, $\Delta^{(t)}=0$  
can be represented in terms of $\Omega$ as,
\begin{eqnarray}
& & 
\begin{array}{l}
\displaystyle{
\kD=-\Ome (\Ome-\oms)^2/2,\ 
\kH=-\Ome (3 \Ome-\oms)/4,\
}\\ \displaystyle{
{z'}^2= - 64 \left(z-(\Ome-\oms)/4\right)^2 \left(z+\Ome/2\right) z.
}
\end{array}
\label{eq-NLS-breather-trig1-tt-paramrep}
\end{eqnarray}


The first degeneracy ($\kt \not=0$),
\begin{eqnarray}
& & {\hskip -12.0truemm}
\left\lbrace
\begin{array}{l}
\displaystyle{  
\kt^2=4 (\Ome-\oms) (3 \Ome-\oms),\ 
p \Kx^2=3 \Ome-\oms,\
}\\ \displaystyle{
\delta e^{i \varphi}=\left(\frac{\Ome-\oms}{2 q}\right)^{1/2}
  e^{\displaystyle{-i \Ome t}} \frac{\sin (\kt t/2)}{\sin(\kt (t-t_3)/2)}\ccomma\
}\\ \displaystyle{
\frac{Q}{\delta}=\frac{\kt}{2(\Ome-\oms)} 
 \left(\cotg(\kt t/2) 
\right. }\\ \displaystyle{ \left. \phantom{12}
  +\frac{6 \Ome-3 \oms + 3 \Ome \cos(\kt t)}
        {3 \Ome[\ax \cosh(\Kx x)+\cos(\kt t/2)]\sin(\kt t/2)} \right),\
}\\ \displaystyle{
\cos(\kt t_3)=-(2\Ome-\oms)/\Ome,\
\sin(\kt t_3)=i \kt/(2 \Ome)\ccomma\
}
\end{array}
\right.
\label{eq-NLS-breather-tt-trig1}
\end{eqnarray}
is the first iterate of the plane wave (\ref{eq-NLS-Qx0})
by the B\"acklund transformation,
it 
depends on two arbitrary real constants $\oms, \Ome$
restricted to $0 < \Ome/\oms < 1$ by the reality of $y(x,t)$.
Depending on the signs of $(\Kx^2,\kt^2)$,
this mathematical solution defines four physical solutions:
two unbounded in the defocusing r\'egime,
and two in the focusing r\'egime:
the Kuznetsov bright soliton solution
\cite{Kuznetsov-NLS} 
\cite[(6.10)]{KI-NLS}
\cite[(41a)]{MaYanchow},
localized in space and periodic in time,
\begin{eqnarray}
& &  {\hskip -2.0truemm}
\left\lbrace
\begin{array}{l}
\displaystyle{
%
%
A=\sqrt{-\Ome/q}  e^{\displaystyle{-i \Ome t}} 
}\\ \displaystyle{
\phantom{1} \times
\left\lbrack 
1-\frac{2(1-\ax^2) \Ome \cos(\kt t/2)+ i (\kt/2) \sin(\kt t/2)}
      {\Ome[\ax \cosh(\Kx x)+\cos(\kt t/2)]} 
\right\rbrack
}\\ \displaystyle{ 
\Kx^2=2 \Ome (1-\ax^2),\
\kt^2=-16 \Ome^2 \ax^2 (1-\ax^2),\
1<\ax^2,\
}
\end{array}
\right.
\label{eq-NLS-Kuznetsov}
\end{eqnarray}
and the breather solution of Akhmediev \cite[Eq.~(11)]{AK}
localized in time and periodic in space,
\begin{eqnarray}
& & {\hskip -2.0truemm}
\left\lbrace
\begin{array}{l}
\displaystyle{
A=\sqrt{-\Ome/q}  e^{\displaystyle{-i \Ome t}} 
}\\ \displaystyle{
\phantom{1} \times
\left\lbrack 
1+\frac{2(\ax^2-1)) \Ome \cosh(\Kt t/2)+ i (\Kt/2) \sinh(\kt t/2)}
      {\Ome[\ax \cos(\iKx x)+\cosh(\Kt t/2)]}   
\right\rbrack
}\\ \displaystyle{ 
\iKx^2=-2 \Ome (1-\ax^2),\
\Kt^2=16 \Ome^2 \ax^2 (1-\ax^2),\
0<\ax^2<1.
}
\end{array}
\right.
\label{eq-NLS-Akhmediev}
\end{eqnarray}
A rigorous proof of their instability under small perturbations
can be found in \cite{AFM-breather-unstable}.


The second degeneracy ($k=0, \oms=3 \Ome \not=0$) yields the Peregrine soliton \cite{Peregrine},
whose complex amplitude is rational in $x$ and $t$,
\begin{eqnarray}
& &  {\hskip -8.0truemm}                                      
A=\left(-\frac{\Ome}{q}\right)^{1/2}
\left\lbrack 
1 + 4 p \frac{1 - 2 i \Ome t}{2 \Ome x^2 - p (1+(2 \Ome t)^2)}
\right\rbrack
 e^{\displaystyle -i \Ome t}. 
\label{eq-NLS-generic-rat}
\label{eq-NLS-breather-tt-trig1-rat}
\end{eqnarray}
whose large maximum amplitude $3$ above its background
makes it a simple prototype of rogue wave.

\subsection{Nongeneric solutions $Q_x z'=0$}


If $Q_x=0$, the solution is a particular plane wave, 
\begin{eqnarray}
& & {\hskip -12.0truemm}
A
 = \sqrt{-\oms/q} e^{\displaystyle -i \oms t},\
\label{eq-NLS-Qx0}
\end{eqnarray}
which is also the limit $\Ome \to \oms$ of  
both (\ref{eq-NLS-Kuznetsov}) and (\ref{eq-NLS-Akhmediev}). 


If $Q_x\not=0$ and $z=z_0\not=0$, one obtains a two-parameter
particular ``dark'' one-soliton solution \cite[(28)]{ZS1973},
\begin{eqnarray}
& & 
%
\begin{array}{l}
\displaystyle{
A=\left(-\frac{2 p}{q}\right)^{1/2}
  \left(\kx \tanh(\kx (x - c t) )+ i \frac{c}{2 p} \right) e^{\displaystyle -i \Omega_0 t},\
}\\ \displaystyle{
\kx^2=\Omega_0/(2 p)-c^2/(4 p^2),\
\Omega_0=\oms + 2 z_0,\
}
\end{array}
\label{eq-NLS-Qxn0-zn0-trigo}
\end{eqnarray}
and its one-parameter rational degeneracy $\kx=0$,
\begin{eqnarray}
& & {\hskip -7.0truemm}
%
A=\left(-\frac{2 p}{q}\right)^{1/2}
  \left(\frac{1}{x-c t} + i \frac{c}{2 p} \right) e^{\displaystyle -i \Omega_0 t},\
c^2=2 p \Omega_0.
\label{eq-NLS-Qxn0-zn0-rat}
\end{eqnarray}


$Q_x\not=0$ and $z=0$ defines the envelope solution, 
\begin{eqnarray}
& & {\hskip -12.0truemm}
\left\lbrace
\begin{array}{l}
\displaystyle{
A=\sqrt{2 p/q} \dn(\kx x,\Jmx)  e^{\displaystyle - i \oms t}
}\\ \displaystyle{
\oms=p \kx^2 (\Jmx-2),\ \kH=p^2 \kx^4 (\Jmx-1),\
}
\end{array}
\right.
\label{eq-NLS-Qxn0-z0-ell}
\end{eqnarray}
and its 
degeneracy 
``bright'' one-soliton solution \cite{ZS1971},
\begin{eqnarray}
& & {\hskip -5.0truemm}
\kH=0:\
A=\left(\frac{2 p}{q}\right)^{1/2}\frac{\kx}{\cosh(\kx x)} e^{\displaystyle -i \oms t},\
\kx^2=-\frac{\oms}{p}.
\label{eq-NLS-Qxn0-z0-sech}
\end{eqnarray}
The other trigonometric degeneracy $\kH=-\oms^2/4$
is identical to the limit $z_0=0$ of (\ref{eq-NLS-Qxn0-zn0-trigo}),
and their common rational degeneracy is 
also the limit $\Omega_0 \to 0$ of (\ref{eq-NLS-Qxn0-zn0-rat}).


\subsection{Nongeneric solutions $Q_x z'\not=0$ and $\kD=0$}

One must distinguish $\kH (\oms^2+4 \kH)$ zero or nonzero.

For $\kH (\oms^2+4 \kH)\not=0$,
one obtains,
\begin{eqnarray}
& & 
\left\lbrace
\begin{array}{l}
\displaystyle{
z^{-1}
           = 2 \at \left(\cos(\kt t) - \cos(\kt \tone)\right),\
					\sin(\kt \tone)=- 4 i/(\at \kt),
}\\ \displaystyle{
y=\partial_t \log \left\lbrack F(x)\sin(\kt t/4) + \cos(\kt t/4)  \right\rbrack,\ 
}\\ \displaystyle{
8 p {F'}^2 =2 \oms (F^2+1)^2 - \at (\kt^2/4) (F^4 - 6 F^2 +1),
}\\ \displaystyle{
\at^2=-\frac{16 \kH}{(\oms^2 + 4 \kH)^2}\ccomma\
\kt^2= 4 (\oms^2 + 4 \kH),\
\cos(\kt \tone)=\frac{8 \oms}{\at \kt^2}\ccomma\
}
\end{array}
\right.
\label{eq-NLS-trigell-detail}
\end{eqnarray}
and the reality of $z(t)$ 
restricts $\kH$ to be negative.
						
To 
our knowledge, 
this is a new solution,
depending on two 
constants $\oms,\kH$.
The reason why it was not found earlier is 
the choice of all 
authors to 
integrate the $x$-elliptic ODE (\ref{systemQfirstintegral})
instead of the $t$-Riccati ODE (\ref{eqsystemQ})${}_1$,
preventing 
$\kD=0$ to be singled out.
%
%
The physically admissible solutions, elliptic in $x$,
exist in focusing and defocusing r\'egimes
but are not bounded.
When $-\oms^2/4 < \kH <0$,
the amplitude $A e^{i \oms t}$ is periodic in time,
\begin{eqnarray}
& & 
\left\lbrace
\begin{array}{l}
\displaystyle{
    A=\left(-\frac{\at}{q \kH}\right)^{1/2} \frac{\kt^2}{16 \sin(\kt \tone/2)}	
			e^{\displaystyle{-i\oms t}}
}\\ \displaystyle{ \phantom{12} \times
\frac{\frac{\at \kt}{4} (\cos(\kt \tone)-1)[1+F(x) c]
      +i [F(x) - c]}
     {F(x) + c}\ccomma
}\\ \displaystyle{
F(x)=c_0 \cs(\kx x,m) \hbox{ real},\
c=\cotg(\kt t/4),\
}
\end{array}
\right.
\label{eq-NLS-trigell}
\label{eq-NLS-trigell-akreal-Jacobi}
\end{eqnarray}
and, when $\kH<-\oms^2/4$, only periodic in $x$,
\begin{eqnarray}
& & 
\left\lbrace
\begin{array}{l}
%
\displaystyle{
    A=\left(\frac{\at}{q \kH}\right)^{1/2} \frac{\kt^2}{16 \sinh(\Kt \tone/2)}	
			e^{\displaystyle{-i\oms t}}
}\\ \displaystyle{ \phantom{12} \times
\frac{\frac{\at \Kt}{4} (\cosh(\Kt \tone)-1)[1+G(x) c]
      +i [G(x) + c]}
        {-G(x) + c}\ccomma\
}\\ \displaystyle{	
G(x)=i F(x) \hbox{ real},\	
c=\coth(\Kt t/4),\
\Kt^2=-\kt^2>0.
}
\end{array}
\right.
\label{eq-NLS-trigell-akmixed-Jacobi}
\end{eqnarray}



The degeneracy $\kD=0$, $\kH =-\oms^2/4\not=0$ 
of (\ref{eq-NLS-trigell}),
\begin{eqnarray}
& & {\hskip -5.0truemm}
A=\sqrt{- \frac{2 p}{q}} \frac{K}{2}\left\lbrack 1- \frac{2(2 \oms t -i)}{\sinh(\sqrt{2 \oms/p} x) + 2 \oms t} \right\rbrack 
e^{\displaystyle - i \oms t},\
\label{eq-NLS-breather-rat1}
\end{eqnarray}
is the limit $\Ome \to \oms$
of the degeneracy (\ref{eq-NLS-Kuznetsov}) of (\ref{eq-NLS-generic-ell}),
obtained by
\begin{eqnarray}
& & 
\begin{array}{l}
\displaystyle{
\Ome=\oms (1 - 2 \varepsilon^2),\
\ax=\varepsilon,\
\kt=-4 i \oms \varepsilon,\ 
}\\ \displaystyle{	
\cosh(\Kx x + i \pi/2)=i \sinh(\kx x),\
\varepsilon \to 0,
}
\end{array}
\end{eqnarray}
and expanding $\sin$ and $\cos$ near $\kt t=0$.
Although we could not find 
(\ref{eq-NLS-breather-rat1})
explicitly written somewhere,
it is certainly not new,
see for instance \cite{ChowKI}.

Last, the degeneracy $\kH=0$ 
has a nonreal value of $z(t)$.

Table \ref{TableAmplitudes} displays all solutions 
generated by (\ref{eqAK-constraint}).


\tabcolsep=0.5truemm

\begin{table}[ht] 
\caption[All solutions]{
         All solutions of the constraint (\ref{eqAK-constraint}).
Each solution is separated by a single line from its degeneracies.
Columns display:
$x$ and $t$-dependences of $A e^{i \oms t}$
(quasi-elliptic Q, elliptic E, trigonometric T, rational R, none $0$),
the arbitrary 
constants,
the complex amplitude,
the initial reference. 
}
\vspace{0.2truecm}
\begin{center}

\begin{tabular}{| c | r | r | r | r | c | c | l | c | l |}
\hline 
   &$Q_x$   & $z'(t)$& $z$    & $\kD$&$x$&$t$& arb      & Eq                          &reference\\ \hline \hline 
$A$&$\not=0$&$\not=0$&$\not=0$&$\not=0$&E&Q&$\oms\kD\kH$&(\ref{eq-NLS-generic-ell})   &\cite[(3), (22), (24)--(25)]{AEK}\\ \hline
$B$&$\not=0$&$\not=0$&$\not=0$&$\not=0$&E&E&$\oms\kD   $&(\ref{eq-NLS-breather-AK})   &\cite[(18)]{AK}\\ \hline
$C$&$\not=0$&$\not=0$&$\not=0$&$\not=0$&T&T&$\Ome\ax   $&(\ref{eq-NLS-Kuznetsov})     &\cite{Kuznetsov-NLS}\\ \hline 
$D$&$\not=0$&$\not=0$&$\not=0$&$\not=0$&T&T&$\Ome\ax   $&(\ref{eq-NLS-Akhmediev})     &\cite[(11)]{AK}\\ \hline
$E$&$\not=0$&$\not=0$&$\not=0$&$\not=0$&R&R&$\oms      $&(\ref{eq-NLS-generic-rat})   &\cite[(6.7)]{Peregrine}\\ \hline 
$1$&$     0$&$      $&$      $&$      $&0&0&$\oms      $&(\ref{eq-NLS-Qx0})           &\cite[(37),(51)]{AEK} \\ \hline \hline 
$2$&$\not=0$&$     0$&$\not=0$&$      $&T&T&$\Omega_0 c$&(\ref{eq-NLS-Qxn0-zn0-trigo})&\cite[(3)]{ZS1973} \\ \hline 
$3$&$\not=0$&$     0$&$\not=0$&$      $&R&R&$\oms      $&(\ref{eq-NLS-Qxn0-zn0-rat})  &\cite{ZS1973} \\ \hline \hline 
$4$&$\not=0$&$     0$&$     0$&$      $&E&0&$\oms\kH   $&(\ref{eq-NLS-Qxn0-z0-ell})   &\cite[(54), (60)]{AEK} \\ \hline 
$5$&$\not=0$&$     0$&$     0$&$      $&T&0&$\oms      $&(\ref{eq-NLS-Qxn0-z0-sech})  &\cite{ZS1971} \cite[(46)]{AEK}\\ \hline\hline
$6$&$\not=0$&$\not=0$&$\not=0$&$     0$&E&T&$\oms\kH   $&(\ref{eq-NLS-trigell})       &New\\ \hline  
$7$&$\not=0$&$\not=0$&$\not=0$&$     0$&T&R&$\oms      $&(\ref{eq-NLS-breather-rat1}) &   \\ \hline \hline \hline 
%
\end{tabular}
\end{center}
\label{TableAmplitudes}
\end{table}


\section{On constraints of higher degree}
\label{sectionHigher-degree-solutions}

Since those singularities of $A$ and $\barA$
which depend on the initial conditions
are simple poles \cite{CCT,CMBook},
the next constraint after (\ref{eqAK-constraint}) should be,
\begin{eqnarray}
& & 
\begin{array}{l}
\displaystyle{
(g_{2,1} \ReA^2+ 2 g_{2,2} \ReA \ImA + g_{2,3} \ImA^2 + g_{2,4} \ReA_x+ g_{2,5} \ImA_x)
}\\ \displaystyle{	
+(g_{1,1} \ReA+g_{1,2} \ImA)+g_0=0,\
\ReA=\Re(A), \ImA=\Im(A),\
}
\end{array}
\label{eq-NLS-constraints}
\end{eqnarray}
in which the real 
coefficients $g_{Nij\dots}$ 
depend on $t$ (and maybe on $x$). 
Indeed,
the relevant degree is the singularity degree (two in (\ref{eq-NLS-constraints})), 
not the polynomial degree,
which is why the restrictive assumption \cite[Eq.~(61)]{AEK}
($g_{2,4}=g_{2,5}=0$)
finds nothing new. 
The larger freedom of (\ref{eq-NLS-constraints})
should generate more solutions,
this will be the subject of future work.


\section{Conclusion and discussion}

The present work, 
which makes explicit the three-parameter extrapolation of the NLS breather,
explains the $t$-period doubling experimentally observed \cite{PhLAM.OL.2020}.
It should provide a much better precision in all the experiments on the 
phenomenon of modulational instability.

The Lam\'e equation is fundamental
in the solution of the constraint (\ref{eqAK-constraint}):
(i) it leads to the compact expression (\ref{eq-NLS-generic-ell}),
(ii) it provides a natural derivation of 
the 
breather (\ref{eq-NLS-breather-AK}),
initially obtained by expert manipulations \cite{AEK,AA-NLS-breather}.

Since the Kuznetsov solution (\ref{eq-NLS-Kuznetsov}),
identical in the complex plane to the Akhmediev breather (\ref{eq-NLS-Akhmediev}),
is generated by the plane wave (\ref{eq-NLS-Qx0})
\textit{via} the B\"acklund transformation (BT),
it is natural to ask which seed generates 
the three-parameter solution (\ref{eq-NLS-generic-ell}),
an extrapolation of (\ref{eq-NLS-Kuznetsov}).
We conjecture that this could be the general traveling wave
\begin{eqnarray}
& & {\hskip -2.0truemm}
A = 
\left(-\frac{2 p}{q} \right)^{1/2}
\frac{\sigma(\xi+d)}{\sigma(\xi)} 
			 e^{\displaystyle{-i \omega t-\zeta(d) \xi + i \frac{c}{2 p}\xi}},
			\xi=x- c t,
\nonumber\\ & &
\label{eqNLS_Traveling_wave_product-of-sigma}
\end{eqnarray}
with $i d$ real (again Lam\'e!)
for two reasons:
(i)
Since the BT involves the integration of a linear differential system
(the Lax pair)
depending on the seed, 
this seed must be elliptic in $x$ and $t$;
(ii) 
The elliptic discriminants 
$\Delta^{(x)}$,
$\Delta^{(t)}$ 
of (\ref{eq-NLS-generic-ell})
have a never zero ratio,
just like the elliptic discriminants of (\ref{eqNLS_Traveling_wave_product-of-sigma})
have for ratio a power of $c$.

\section*{Acknowledgments}

The author is pleased to thank Micheline Musette for 
a critical reading of the manuscript.
This work was initiated in 
Centre international de rencontres math\'ematiques, Marseille
(grant 2311, year 2019),
whose hospitality is gratefully acknowledged.

\section*{Conflict of interest}

The author declares that he has no conflicts of interest.

\vfill \eject 

\end{document}